\documentclass[lettersize,journal]{IEEEtran}
\usepackage{amsmath,amsfonts}
\usepackage{algorithm2e}
\usepackage{array}
\usepackage[caption=false,font=normalsize,labelfont=sf,textfont=sf]{subfig}
\usepackage{textcomp}
\usepackage{stfloats}
\usepackage{url}
\usepackage{verbatim}
\usepackage{graphicx}
\usepackage{cite}
\usepackage{float}
\usepackage{mathtools}
\usepackage{graphicx} 
\usepackage{amsmath}
\usepackage{amsmath}
\usepackage{amsmath, amssymb}
\usepackage{xcolor}	

\usepackage{algpseudocode}
\usepackage{graphics}
\usepackage{amsmath, amsthm}

\usepackage[caption=false,font=normalsize,labelfont=sf,textfont=sf]{subfig}

\newcounter{assumption}

\newcounter{theorm} 

\begin{document}
	
	\title{Fluid Antenna-Enabled Hybrid NOMA and AirFL Networks Under Imperfect CSI and SIC
	}
	
	
	
	\author{IEEE Publication Technology,~\IEEEmembership{Staff,~IEEE,}
	}
	
	

	\author{Saeid Pakravan, Mohsen Ahmadzadeh, Ming Zeng, Ghosheh Abed Hodtani and Xingwang Li
		
		\thanks{S. Pakravan and M. Zeng are with the Department of Electric and Computer Engineering, Laval University, Quebec City, CA. email: saeid.pakravan.1@ulaval.ca; ming.zeng@gel.ulaval.ca.}
		
		
		\thanks{M. Ahmadzadeh and G. Abed Hodtani are with the Department of Electric and Computer Engineering, Ferdowsi University, Mashhad, Iran. email: m.ahmadzadehbolghan@mail.um.ac.ir; hodtani@um.ac.ir.}

        \thanks{Xingwang Li is with the School of Physics and Electronic Information Engineering, Henan Polytechnic University, Jiaozuo, China. email: lixingwang@hpu.edu.cn.}

		
	}

	\maketitle

	\begin{abstract}
		
		The integration of communication and computation is essential for next-generation wireless systems, especially in scenarios demanding massive connectivity and ultra-low latency. Over-the-air federated learning (AirFL), leveraging the superposition nature of wireless channels, enables fast data aggregation, while non-orthogonal multiple access (NOMA) offers spectrum-efficient connectivity. This paper investigates a fluid antenna (FA)-aided hybrid network, supporting hybrid users comprising both AirFL and NOMA participants. The dynamic reconfigurability of FAs offers significant potential for mitigating interference and enhancing network performance by adapting antenna positions in response to changing channel conditions. We consider practical challenges arising from imperfect channel state information (CSI) and residual interference due to imperfect successive interference cancellation (SIC). To jointly evaluate the learning and communication performance, a hybrid rate metric is introduced. Subsequently, we formulate a robust optimization problem that jointly minimizes the aggregation error while ensuring reliable user communication under CSI and SIC uncertainties. This joint optimization is formulated as a non-convex problem, complicated by the intricate interactions between NOMA and AirFL users and the impact of imperfect CSI and SIC. To solve this problem effectively, we reformulate the optimization as a Markov decision process and solve it using a long short-term memory deep deterministic policy gradient (LSTM-DDPG) algorithm, a memory-based approach within the realm of deep reinforcement learning. Simulation results demonstrate the superiority of the proposed FA-assisted approach over fixed-antenna baselines, particularly under imperfect CSI and SIC conditions, in terms of hybrid rate performance.
		
	\end{abstract}
	
	
	\section{Introduction}

The increasing proliferation of intelligent devices and latency-sensitive applications in modern wireless networks has necessitated a paradigm shift from conventional communication-centric designs to architectures that seamlessly integrate both communication and computation \cite{9210812, 10812728}. This transition is driven by the demands of next-generation wireless networks, which are expected to support massive device connectivity, ultra-low latency, and diverse quality-of-service (QoS) requirements, all while operating under stringent spectral and energy constraints \cite{8808168}. 

Federated learning (FL) has emerged as a promising decentralized machine learning paradigm that allows collaborative model training across edge devices without exchanging raw data \cite{44444}. FL preserves data privacy and mitigates network congestion; however, conventional FL implementations often rely on digital transmission with orthogonal resource allocation, which can introduce high latency and bandwidth inefficiencies in networks with large numbers of participants \cite{9141214}. To alleviate these limitations, over-the-air federated learning (AirFL) has been proposed, leveraging the superposition property of wireless multiple-access channels (MACs) to perform analog model aggregation directly over the air \cite{8952884}. AirFL offers notable reductions in communication latency and improved scalability, making it particularly attractive for real-time edge intelligence applications \cite{9562487, 10400441}.

In parallel, non-orthogonal multiple access (NOMA) has been extensively investigated as a means to enhance spectral efficiency by enabling multiple users to transmit simultaneously over shared resources through power-domain multiplexing \cite{8352616, 77777, Lin2025}. By allocating distinct power levels to users, NOMA increases system capacity without consuming additional bandwidth. Recent studies have investigated NOMA in combination with other emerging technologies, showing substantial gains in both spectral efficiency and user fairness \cite{10912473, 11106811, 10044972, 8642812}.

The integration of AirFL and NOMA introduces a hybrid AirFL-NOMA framework, where hybrid refers to the simultaneous coexistence of two types of users: AirFL users, which perform analog model aggregation, and NOMA users, which transmit conventional data traffic over shared wireless resources. This joint design enables communication- and learning-oriented transmissions to share the same spectrum, thereby enhancing spectral efficiency and supporting large-scale, low-latency, and bandwidth-efficient distributed learning. However, this integration also gives rise to several key challenges. The coexistence of AirFL and NOMA users leads to complex interference patterns and heterogeneous QoS requirements, as well as performance degradation caused by practical imperfections such as imperfect channel state information (CSI) and residual errors from successive interference cancellation (SIC). These effects—often overlooked in theoretical analyses—can significantly degrade both learning accuracy and communication reliability, ultimately limiting the practical effectiveness of hybrid AirFL-NOMA systems.

	To mitigate these challenges, this paper proposes a hybrid AirFL-NOMA framework assisted by fluid antennas (FAs). FAs provide a reconfigurable and flexible architecture that enables dynamic adjustment of antenna positions within a predefined spatial region \cite{10599127, 10528324, pakravan2026fluid1, 11111711, 9264694, saeidpwcnc}. This adaptability introduces a new form of spatial diversity, which can be leveraged to suppress interference and enhance signal quality. By exploiting the spatial adaptability of FAs, the proposed framework enhances both the model aggregation performance of AirFL users and the decoding reliability of NOMA users, particularly under dynamic and uncertain wireless conditions. Compared with prior FA-assisted AirFL and FA-NOMA studies \cite{11106811,11111711,saeidpwcnc,saeidGlobecom}, our framework introduces several key distinctions. Most existing works focus exclusively on FA-assisted NOMA or FA-enabled AirFL, without jointly addressing the coexistence of both user types or the impact of practical imperfections. For instance, \cite{11106811} investigates FA-assisted uplink NOMA under imperfect SIC, but does not incorporate AirFL aggregation. The works in \cite{11111711} and \cite{saeidpwcnc} employ AI-driven FA designs for client selection and improved AirFL aggregation. \cite{saeidGlobecom} studies robust resource allocation for FA-assisted over-the-air computation. In contrast, we simultaneously optimize FA positions, AP beamforming, and transmit power allocation for both AirFL and NOMA users, yielding a unified and practical solution for next-generation wireless systems.

	A key innovation of this work is the introduction of a hybrid rate metric that jointly quantifies the performance of communication and learning tasks in the presence of CSI and SIC imperfections. We formulate a robust optimization problem that aims to maximize the achievable hybrid rate by jointly optimizing the positions of the FAs, the beamforming vector at the access point (AP), and the transmit power allocation for each user, while satisfying the minimum rate requirements of NOMA users and the maximum aggregation error of AirFL users. 
    
    Due to the non-convex nature of the problem and the strong coupling between AirFL and NOMA users, conventional optimization methods are insufficient. To overcome this, we reformulate the problem as a Markov decision process (MDP) and propose a memory-based deep reinforcement learning (DRL) approach using the long short-term memory deep deterministic policy gradient (LSTM-DDPG) algorithm, which
leverages temporal correlations in wireless environments for
adaptive and sequential decision-making.

The key contributions of this paper are summarized as follows:
	\begin{itemize}
		\item We propose a hybrid AirFL-NOMA system architecture enhanced by FAs to support efficient and adaptive integration of learning and communication tasks.
		\item We account for practical imperfections, including CSI uncertainty and residual SIC error, which significantly affect the system performance in realistic scenarios.
		\item We introduce a hybrid rate metric that captures the trade-offs between learning accuracy and communication reliability in heterogeneous user environments.
		\item We formulate a robust joint optimization problem and develop a memory-aware DRL solution based on LSTM-DDPG to achieve near-optimal performance under dynamic conditions.
		\item Simulation results demonstrate that the proposed FA-assisted framework significantly outperforms conventional fixed-antenna baselines, particularly in the presence of CSI and SIC uncertainties.
	\end{itemize}
	
	The rest of the paper is structured as follows. Section~II introduces the system model, describing the overall network architecture, underlying assumptions, and key parameters. Section~III formulates the optimization problem, including the objective function and associated constraints. Section~IV details the proposed LSTM-DDPG-based solution methodology. Section~V provides simulation results and offers a comprehensive performance evaluation of the proposed scheme. Finally, Section~VI concludes the paper.

	{\bf{Notation:}} Italic letters represent scalar quantities, while boldface letters denote vectors. The transpose and conjugate transpose operations are represented by $(\cdot)^T$ and $(\cdot)^H$, respectively. The expectation operator is denoted by $\mathbb{E}[\cdot]$. The real part of a complex number is indicated by $\Re\{\cdot\}$, the symbol $|\cdot|$ represents either the absolute value of a scalar or the cardinality of a set, and $\|\cdot\|$ denotes the Euclidean norm. The notation $\hat{\cdot}$ is used to indicate an estimated value.

	\section{System Model}

	\begin{figure}[]
		\includegraphics[width=8.5cm, height=4.5cm]{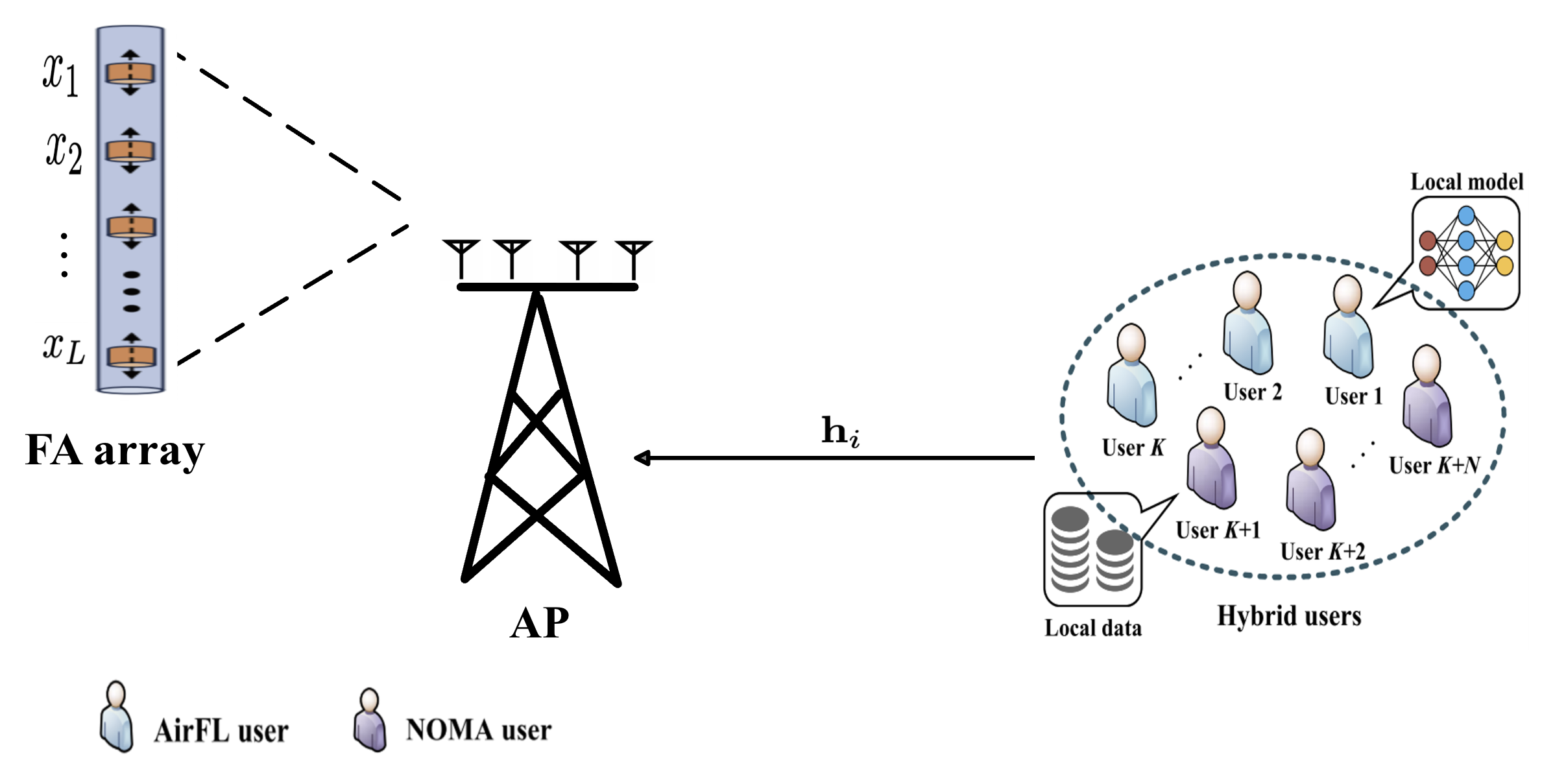}
		\caption{System model of an FA-aided hybrid AirFL-NOMA network.
		}
		\label{fig:NUMRIS}
	\end{figure}
	
	\subsection{Network and Channel Model}

	As depicted in Fig. 1, we investigate a wireless network consists of an AP serving hybrid users, which are classified into NOMA and AirFL users. Both user types employ non-orthogonal communication schemes over a shared wireless MAC. The transmission occurs over $T$ time slots indexed by $t\in \left\{ 1,...,T \right\}$. The hybrid users are represented by the set \(\mathcal{I} = \{1, 2, \ldots, K, K + 1, K + 2, \ldots, K + N\}\), where AirFL users are indexed by \(\mathcal{K} = \{1, 2, \ldots, K\}\) and NOMA users by \(\mathcal{N} = \{K + 1, K + 2, \ldots, K + N\}\). Furthermore, the AP is equipped with $L$ FAs arranged in a linear array, which can dynamically reposition themselves along a one-dimensional line segment of length \(X\). The position of each FA is constrained within the interval $[0, X]$, with a minimum distance $X_0$ maintained between adjacent FAs to avoid antenna coupling. The positions of the \(L\) FAs are collectively denoted by the vector $\mathbf{x} = [x_1, x_2, \ldots, x_L]^T$, where their arrangement satisfies $x_1 < x_2 < \cdots < x_L$. This flexible positioning of the FAs allows the system to adaptively optimize the spatial configuration in response to varying channel conditions, thereby enhancing signal reception and overall system performance. 

	
	The complex channel gain vector between the $i$-th user and the AP, denoted by $\mathbf{h}_i[\mathbf{x}] \in \mathbb{C}^{L \times 1}$, follows a Rician fading model as:
	\begin{equation}
		\mathbf{h}_i[\mathbf{x}] = \sqrt{\frac{A_L d_i^{-\alpha_L} \kappa_r}{\kappa_r + 1}} \mathbf{h}_i^{\text{LOS}}[\mathbf{x}] + \sqrt{\frac{A_N d_i^{-\alpha_N}}{\kappa_r + 1}} \mathbf{h}_i^{\text{NLOS}}, \ \forall i\in \mathcal{I},
	\end{equation}
	where $\kappa_r$ represents the Rician factor, \(d_i\)
	is the distance between the FAs and the $i$-th user, and $A_L$ and $A_N$ are the path loss at the reference distance for the line-of-sight (LoS) and non-line-of-sight (NLoS) components, respectively. The parameters $\alpha_L$ and $\alpha_N$ denote the path loss exponents for the LoS and NLoS components, respectively. The term $\mathbf{h}_i^{\text{LOS}}[\mathbf{x}]$ represents the LoS component, while $\mathbf{h}_i^{\text{NLOS}}$ denotes the NLoS component. All $\mathbf{h}_i^{\text{NLOS}} \in \mathbb{C}^{L \times 1}$ follow an i.i.d. complex Gaussian distribution with zero mean and unit variance. The LoS component $\mathbf{h}_i^{\text{LOS}}[\mathbf{x}]$ can be
	calculated as follows \cite{saeidpwcnc,
		10694601, saeidGlobecom}:
	\begin{equation}
		\mathbf{h}^{\text{LOS}}_{i}[\mathbf{x}] =  [e^{j \frac{2\pi}{\lambda} x_1 \cos(\phi_{i})}, \ldots, e^{j \frac{2\pi}{\lambda} x_L \cos(\phi_{i})} ]^T,
	\end{equation}
	where \(\lambda\) and \(\phi_i\) are the wavelength and the angle of arrival (AoA) of the LoS path, respectively, determined by the positions of the users. It is assumed that each user moves within a predefined area and subsequently transmits information from a stationary position. Furthermore, since the signal path length is considerably greater than the movement range of the FAs, the far-field condition between the AP and the user devices is assumed to hold. As a result, the parameters \( \phi_i \) and \( d_i \) vary randomly across time slots but remain constant during each transmission phase. Thus, during transmission, \( \phi_i \) and \( d_i \) are treated as constants, independent of any changes in the FA positions.

	\subsection{FL Training Process}
	
	This subsection provides a detailed overview of the training procedure in FL, outlining the sequential steps involved in each training round or communication iteration denoted by $\mathcal{T} = \{1, 2, \ldots, T\}$.
	
	\begin{enumerate}
		\item \textbf{Global Model Broadcast:} 
		At the beginning of each training round, the AP initiates the process by broadcasting the latest version of the global model, denoted as $w^{(t)}$. This model is shared with all AirFL users, where $w^{(t)} \in \mathbb{R}^Q$, and $Q$ indicates the dimensionality of the parameter space of the global model. This step ensures that all AirFL users start the training round with the same global model, providing a common reference point for subsequent updates.

		\item \textbf{Local Model Update:} Subsequently, each AirFL user independently updates its local gradient $g_k^{(t)}$ based on its individual dataset. Specifically, the local gradient of the $k$-th user is obtained as:
		\begin{equation}
			g_k^{(t)} = \nabla F_k(w^{(t)}) = \frac{1}{|\mathcal{D}_k|} \sum_{i=1}^{|\mathcal{D}_k|} \nabla f(w_{k}^{(t)}, x_{k,i},y_{k,i}), 
		\end{equation}
		where $\nabla f(w_{k}^{(t)}, x_{k,i},y_{k,i})$ is the gradient of the local loss function $f(w_{k}^{(t)}, x_{k,i},y_{k,i})$ for the corresponding local training sample $(x_{k,i},y_{k,i}), i \in \{1, . . . ,|\mathcal{D}_k|\}$ at client $k$. It is assumed that all local datasets have the same size, denoted as $|\mathcal{D}_k| = D$. 
		
		\item \textbf{Model Aggregation:} 
		Following the computation of local gradients, all AirFL users transmit their respective gradients $\{g_k^{(t)}\}$ to the AP via wireless channels for global synchronization. Upon receiving the local gradients from all AirFL users, the AP updates the global model using the following equation:
		\begin{equation}
			w^{(t+1)} = w^{(t)} - \lambda g^{(t)}, \quad \text{with} \quad g^{(t)} = \frac{1}{K} \sum_{k \in \mathcal{K}} g_k^{(t)},
		\end{equation}
		where $g^{(t)}$ represents the global gradient computed as the average of all local gradients. Here, $\lambda$ denotes the learning rate chosen by the AP.
	\end{enumerate}

	The training process continues iteratively, with each round repeating these steps until either the convergence criteria are satisfied or the predefined maximum number of iterations, denoted as $T$, is reached.

	\subsection{FA-Assisted Concurrent Transmission}
	
	In the $t$-th training round of the FA-assisted communication system, simultaneous uplink transmission is enabled by concurrently encoding and transmitting heterogeneous data from different user types. Specifically, the $n$-th NOMA user encodes its local data $d_n^{(t)}$ into communication symbols $s_n^{(t)}$, while the $k$-th AirFL user encodes its local model gradients $g_k^{(t)}$ into computation symbols $s_k^{(t)}$. This encoding process ensures that both communication and computation data are appropriately formatted for wireless transmission. Subsequently, both NOMA and AirFL users transmit their respective symbols concurrently over the same time-frequency resources. Consequently, the received superposition signal at the AP is expressed as:
	\begin{multline}
		\boldsymbol{y}^{(t)} =  \underbrace{\sum_{k=1}^{K} 
			\boldsymbol{w}^H        
			{\boldsymbol{h}}_k^{(t)}[\mathbf{x}] \sqrt{p_k^{(t)}} s_k^{(t)}}_{\text{AirFL users}} \\+ \underbrace{\sum_{n=K+1}^{K+N} \boldsymbol{w}^H        
			{\boldsymbol{h}}_n^{(t)}[\mathbf{x}]\sqrt{p_n^{(t)}} s_n^{(t)}}_{\text{NOMA users}} + \underbrace{\boldsymbol{w}^H\boldsymbol{z}_0^{(t)}}_{\text{Noise}},
	\end{multline}
	where $p_k^{(t)}$ ($p_n^{(t)}$) denotes the power scaling factor at the $k$-th ($n$-th) user, \( \boldsymbol{w}\in \mathbb{C}^{L \times 1} \) is the receive beamforming vector, $\boldsymbol{z}_0^{(t)} \sim \mathcal{CN}(0, \sigma^2)$ represents additive white Gaussian noise (AWGN) with zero mean and noise power $\sigma^2$.
	
	To ensure efficient power control, it is assumed that the information symbols of all users follow a zero-mean Gaussian distribution with unit variance, and are statistically independent and identically distributed (i.i.d.), i.e., for all $i$, $\mathbb{E}(|s_i^{(t)}|) = 0$, $\mathbb{E}(|s_i^{(t)}|^2) = 1$, and $\mathbb{E}((s_i^{(t)})^H s_j^{(t)}) = 0$, $\forall i \neq j$. Consequently, the transmit power constraint for the $i$-th user is:
	\begin{equation}
		\mathbb{E}(|\sqrt{p_i^{(t)}} s_i^{(t)}|^2) = |p_i^{(t)} |\leq P_i^\text{max}, \quad \forall i \in \mathcal{I},
	\end{equation}
	where $P_i^\text{max} > 0$ represents the maximum transmit power.

	Despite the advantages of concurrent transmission in terms of spectral efficiency, practical wireless systems are subject to various imperfections that can degrade overall performance. Two key impairments are imperfect CSI and suboptimal SIC at the receiver.
	To ensure robust system design, it is imperative to incorporate both CSI and SIC imperfections in the analytical framework. This facilitates accurate performance evaluation under real-world conditions and supports the development of resilient signal processing and resource allocation strategies.

	\textbf{Imperfect CSI.} In real-world deployments, the estimated channel information at the receiver often deviates from the actual channel due to noise, interference, and hardware limitations \cite{9495269, 10261509}. To model this, the imperfect channel between the $i$-th user and the AP is represented as
	\begin{equation}
		\boldsymbol{h}_i = \hat{\boldsymbol{h}}_i + \boldsymbol{\varepsilon}_{h,i}, \quad \forall i \in \mathcal{I},
	\end{equation}
	where $\hat{\boldsymbol{h}}_i$ denotes the estimated channel, and $\boldsymbol{\varepsilon}_{h,i} \sim \mathcal{CN}(0, \sigma_{h,i}^2)$ captures the channel estimation error, modeled as a circularly symmetric complex Gaussian (CSCG) random variable with variance $\sigma_{h,i}^2$. This model allows for a realistic evaluation of the system’s robustness under CSI uncertainty and supports the design of resilient transmission and resource allocation strategies.

	{\bf{Imperfect SIC.}} In the proposed hybrid system, SIC is employed at the AP to mitigate interference by sequentially decoding the signals of strong users (i.e., NOMA users) before extracting the weak users’ signals (i.e., AirFL users). However, in practical wireless communication systems, imperfect SIC at the receiver may arise due to hardware constraints in decoding and canceling interfering signals. Despite the advancement of signal processing techniques, residual interference from strong users can persist, degrading system capacity and reliability. To address these challenges, it is essential to incorporate imperfect SIC models into channel frameworks. Such integration enables a thorough evaluation of system resilience and performance degradation under real-world conditions. The residual interference in imperfect SIC scenarios is commonly modeled using a scalar parameter $\epsilon_b \in [0,1]$, which captures the fraction of interference power remaining after cancellation. This approach has been adopted in several prior studies, such as \cite{Kwang2023, XLi2020, deSena2020}, to reflect the practical limitations of SIC in realistic NOMA deployments. Specifically, $\epsilon_b = 0$ corresponds to ideal SIC with no residual interference, whereas $\epsilon_b = 1$ represents the worst-case scenario where interference is fully retained. Employing SIC techniques, the AP sequentially decodes signals from strong users to mitigate interference, subsequently extracting the superposed signal of weak users. Under this scenario, the SIC decoding order at the AP for all users can be determined as follows:
	\begin{equation}
		\underbrace{|\hat{\boldsymbol{h}}_k^{(t)}|^{2} \leq |{\hat{\boldsymbol{h}}}_{k+1}^{(t)}|^{2} \leq \ldots \leq }_{\substack{\text{weak signals} \\ \text{that cannot be canceled}}}|{\hat{\boldsymbol{h}}}_{n}^{(t)}|^{2} \underbrace{\leq \ldots \leq |{\hat{\boldsymbol{h}}}_{K+N}^{(t)}|^{2}}_{\substack{\text{strong signals} \\ \text{that can be partially canceled}}}. \end{equation}

	Accordingly, the signal-to-interference-plus-noise ratio (SINR) for the $n$-th NOMA user at the AP under imperfect SIC is expressed as: 
	\begin{multline}
		\gamma_n^{(t)} =\\ \frac{p_n^{(t)} |\boldsymbol{w}^H{\boldsymbol{h}}_n^{(t)}|^2}{\sum_{k=1}^{n-1} p_k^{(t)} |\boldsymbol{w}^H{\boldsymbol{h}}_k^{(t)}|^2 + \epsilon_b \sum_{i=n+1}^{N+K} p_i^{(t)} |\boldsymbol{w}^H{\boldsymbol{h}}_i^{(t)}|^2 + {\left\|\boldsymbol{w}\right\|^2}\sigma^2}.
	\end{multline}
Notably, the computation symbols from the weaker AirFL users remain uncanceled and thus contribute as interference to the stronger NOMA users.

	Following Shannon's capacity theorem, the achievable uplink data rate for the $n$-th NOMA user in the $t$-th communication round is given by
	\begin{equation}
		R_n^{(t)} =B \log_2 (1 + \gamma_n^{(t)}),
	\end{equation}
	where $B$ denotes the bandwidth available at the AP. Thus, the overall uplink sum rate of the NOMA users can be calculated as:
	\begin{equation}
		R_{\text{NOMA}}^{(t)} = \sum_{n=K+1}^{K+N} R_n^{(t)}.
	\end{equation}
	
	Furthermore, due to the imperfections in CSI and SIC, the residual signal used in AirFL model aggregation can be represented as follows: 
	\begin{multline}
		\hat{\boldsymbol{y}}^{(t)} = \sum_{k=1}^{K} \boldsymbol{w}^H(\hat{\boldsymbol{h}}_k^{(t)}+\boldsymbol{\varepsilon}_{h,k}^{(t)}
		) \sqrt{p_k^{(t)}} s_k^{(t)} \\+ \sqrt{\epsilon_b} \sum_{n=K+1}^{K+N} \boldsymbol{w}^H(\hat{\boldsymbol{h}}_n^{(t)}+\boldsymbol{\varepsilon}_{h,n}^{(t)}
		) \sqrt{p_n^{(t)}} s_n^{(t)} + \boldsymbol{w}^H\boldsymbol{z}_0^{(t)}.
	\end{multline}
The recovered signal for AirFL users is then given by $\hat{s}^{(t)} = \frac{\hat{\boldsymbol{y}}^{(t)}}{K}$, which is compared against the ideal aggregated result $s^{(t)} = \frac{1}{K} \sum_{k=1}^{K} s_k^{(t)}$. The computation distortion of the recovered average gradient message with respect to the ideal average message $s^{(t)}$ under imperfect CSI and SIC scenario, is characterized by the instantaneous mean squared error (MSE) between $\hat{s}^{(t)}$ and $s^{(t)}$, defined as:
	\begin{multline}
		{\text{MSE}(\hat{s}^{(t)}, s^{(t)}) = \mathbb{E}(|\hat{s}^{(t)} - s^{(t)}|^2)} \\=
		\underbrace{\frac{1}{K^2} \sum_{k=1}^{K} \bigg|\boldsymbol{w}^H\hat{\boldsymbol{h}}_k^{(t)} \sqrt{p_k^{(t)}} - 1\bigg|^2 }_{\text{Signal misalignment error}}+ \underbrace{\frac{\epsilon_b }{ K^2} \sum_{n=K+1}^{K+N} p_n^{(t)}|\boldsymbol{w}^H\hat{\boldsymbol{h}}_n^{(t)} |^2}_{\text{SIC-related error}} \\+\underbrace{\frac{1 }{K^2} \sum_{k=1}^{K} p_k^{(t)} {\left\|\boldsymbol{w} \right\|^2}\sigma^{2}_{h,k}}_{\text{CSI-related error}}+\underbrace{\frac{\epsilon_b }{K^2\eta} \sum_{n=K+1}^{K+N} p_n^{(t)}{\left\|\boldsymbol{w} \right\|^2}\sigma^{2}_{h,n} }_{\text{SIC and CSI-related error}}\\+\underbrace{\frac{ {\left\|\boldsymbol{w} \right\|^2}\sigma^2}{K^2},}_{\text{Noise-induced error}} 
	\end{multline}
	where the expectation is taken over the randomness of $s_i^{(t)}$, $\boldsymbol{\varepsilon}_{h,i}^{(t)}$, and $z_0^{(t)}$. 

	Given that the MSE serves as a comprehensive metric for quantifying computation distortion across all AirFL users, we now proceed to define the system’s collaborative computation performance in terms of its achievable computation rate. Specifically, the achievable rate for global model aggregation at iteration $t$ is defined as \cite{10092857, 9798757}:
	\begin{equation}
		R_{\mathrm{AirFL}}^{(t)} = B \log_2 \left( 1 + \frac{\mathbb{E}(|\hat{s}^{(t)}|^2) - \mathrm{MSE}(\hat{s}^{(t)}, s^{(t)})}{\mathrm{MSE}(\hat{s}^{(t)}, s^{(t)})} \right),
		\label{eq:RAirFL}
	\end{equation}
	where $\mathbb{E}(|\hat{s}^{(t)}|^2)$ represents the total received power of the reconstructed signal.
	
	It is worth noting that the computation rate in (14) resembles the logarithmic form of the communication rate, thereby enabling a unified framework for evaluating and comparing the performance of heterogeneous users. Leveraging the rate expressions derived for both NOMA users and AirFL users, the overall achievable hybrid computation-communication rate for the integrated system is expressed as
	\begin{equation}
		\label{hybrid_rate}
		R_{\mathrm{Hybrid}}^{(t)} = (1 - \lambda) R_{\mathrm{NOMA}}^{(t)} + \lambda R_{\mathrm{AirFL}}^{(t)},
	\end{equation}
	where $\lambda \in [0, 1]$ denotes a weighting factor that balances the performance contribution between NOMA and AirFL users. The choice of $\lambda$ plays a critical role in determining the resource allocation strategy across the two user types. Specifically, $\lambda$ adjusts the trade-off between prioritizing communication performance (NOMA users) and computation performance (AirFL users). A larger \(\lambda\) emphasizes AirFL aggregation accuracy and accelerates model convergence, whereas a smaller \(\lambda\) prioritizes the data rates of NOMA users, enhancing conventional communication throughput.

	\section{Problem Formulation}

	This paper focuses on maximizing the hybrid computation-communication rate in a FA-assisted uplink hybrid network. Specifically, we jointly optimize three key components: the transmit power $\boldsymbol{p}$ at the users, the receive beamforming vector, and the FA positions. The optimization problem is formulated as follows: 
	\begin{equation}
		\begin{aligned}
			\mathcal{OP}: &\max_{ \boldsymbol{p}, 
				\boldsymbol{w}, \mathbf{x}} \ (1 - \lambda) R_{\mathrm{NOMA}}^{(t)} + \lambda R_{\mathrm{AirFL}}^{(t)}, \\
			\text{s.t.} 
			&\quad C_1: B \log_2 (1 + \gamma_n^{(t)})\ge R_n^{\min}, \quad \forall n\in \mathcal{N}, \\
			& \quad C_2: \mathrm{MSE}(\hat{s}^{(t)}, s^{(t)}) \leq \varepsilon_0,
			\\
			&\quad C_3: 0 \leq x_l \leq X, \quad \forall l \in \{1, \ldots, L\}, \\
			& \quad C_4: x_1 < x_2 < \ldots < x_L, \\
			& \quad C_5: x_l - x_{l-1} > X_0, \quad \forall l \in \{2, \ldots, L\},\\
			& \quad C_6: (6) \ \text{and} \ (8).\\
		\end{aligned}
	\end{equation}
	Here, $R_n^{\min}$ denotes the minimum required communication rate for the $n$-th NOMA user, and $\varepsilon_0 > 0$ defines the maximum allowable MSE for the aggregation process in AirFL.
	
	Constraint $C_1$ guarantees the QoS for NOMA users by ensuring a minimum data rate. Constraint $C_2$ imposes a limit on the global aggregation error to preserve AirFL accuracy. The threshold $\varepsilon_0$ is selected to ensure a balance between AirFL model convergence and network resource efficiency, where smaller values improve aggregation accuracy and accelerate learning convergence, while larger values reduce communication overhead and allow more flexible FA positioning.
    Constraints $C_3$-$C_5$ impose physical limitations on the FA positions: $C_3$ bounds each FA's location within a predefined region $[0, X]$; $C_4$ ensures a strictly increasing order of antenna positions to avoid overlap; and $C_5$ guarantees a minimum spacing $X_0$ between adjacent antennas to mitigate inter-antenna interference. Finally, constraint $C_6$ refers to additional conditions defined earlier in (6) and (8), which pertain to the users' maximum transmit power and the SIC decoding order, respectively.

	The formulated optimization problem is inherently non-convex due to the strong coupling between variables, resulting in complex constraints and a non-concave objective function, making traditional convex optimization techniques inefficient for finding optimal solutions. Moreover, conventional methods are unsuitable for real-time decision-making in dynamic environments. To address these challenges, we reformulate the problem as a MDP and propose a DRL-based solution. This algorithm continuously interacts with the environment, generating optimal actions by maximizing a reward function while incorporating constraints through penalty mechanisms, thereby ensuring dynamic and efficient real-time adaptation.

\section{Proposed DRL Algorithm} 
	
To address this optimization problem, we propose a DRL framework, wherein an intelligent agent is deployed at the AP. The agent is designed to autonomously learn an optimal policy for jointly optimizing the beamforming vector $\boldsymbol{w}$, the power allocation vector $\boldsymbol{p}$, and the FA location vector $\boldsymbol{x}$ during each training round, with the objective of maximizing the hybrid communication rate. Details of the MDP are provided below:
	\begin{itemize}
		\item \textbf{State Space}: The state at time step $t$, denoted by $s_t$, comprises the set of distances $d_I$ between the FAs and the users, as well as the AoA of the LoS paths $\phi_I$, for all $i \in \mathcal{I}$. This can be expressed as: $s_t = [[d_{1,t}, \ldots, d_{I, t}], [\phi_{1,t}, \ldots, \phi_{I,t}]]$.
		\item \textbf{Action Space}:The state at time step $t$, denoted as $a_t$, consists of the beamforming vector, the locations of the FAs, and the transmit power vector. Therefore, the action space is expressed as: $a_t = [[w_{1,t}, \ldots, w_{L,t}], [x_{1,t}, \ldots, x_{L,t}], [p_{1,t}, \dots, p_{{K+N},t}]].$
		\item \textbf{Reward function}:
		Based on (\ref{hybrid_rate}), in order to maximize the hybrid rate, the reward function is formulated as:
\begin{equation}
	r(s_t, a_t) = 
	\begin{cases}
		r_p, & \mathclap{\text{if } |\boldsymbol{w}^{H} \boldsymbol{h}_{n}^{(t)}[\mathbf{x}]| = 0} \\
		(1 - \lambda) R_{\mathrm{NOMA}}^{(t)} + \lambda R_{\mathrm{AirFL}}^{(t)}, & \text{otherwise}
	\end{cases}
\end{equation}
		where \( r_p \) is a penalty term that needs to be tuned during the simulation to achieve optimal convergence behavior. The penalty term \(r_p\) is designed to discourage infeasible actions, such as beamforming vectors leading to zero effective channel gains. Its value directly affects the learning performance: if \(|r_p|\) is too small, the agent may explore invalid actions excessively, slowing convergence; if \(|r_p|\) is too large, the agent may become overly conservative, limiting exploration and potentially yielding suboptimal policies. In practice, \(r_p\) is tuned empirically to balance exploration and convergence speed.
	\end{itemize}

Traditional model-free value-based DRL algorithms, such as deep Q-networks (DQN), are designed for discrete action spaces and cannot effectively handle continuous control problems. To address this, we employ policy gradient-based reinforcement learning methods. Specifically, the DDPG algorithm is an off-policy actor-critic approach that is well-suited for handling continuous action spaces. However, the standard DDPG framework, which relies on fully connected deep neural networks (DNNs), is inadequate for modeling the temporal dynamics present in environments such as user mobility \cite{111111, 222222}. To overcome this limitation, we enhance the DDPG framework by integrating LSTM networks, enabling the model to effectively capture temporal dependencies and continuously adapt to the changing environment. In our implementation, the proposed LSTM-DDPG framework employs four interconnected neural networks: an actor network, denoted as \(\pi_\phi\) with parameters \(\phi\),  which selects actions \(a_t = \pi_\phi(s_t) + \xi\) based on the states \(s_t\), where \(\xi\) is a noise term introduced for exploration; a critic network, represented as \(Q_\theta\), with parameters \(\theta\), which evaluates state-action pairs by computing their corresponding Q-values \(Q_\theta(s_t, a_t; \theta)\); a target actor network, which is a copy of the actor network with updated parameters; and a target critic network, which is a copy of the critic network.

	The LSTM-DDPG framework aims to find an optimal policy, denoted as  \(\pi^*\), that maximizes cumulative returns, defined as the sum of expected future rewards, which can be formulated as:
	\begin{equation}
		\pi^* = \arg \max_{\pi} \mathbb{E}_{s_t, a_t} \left[ \sum_{t=0}^{\infty} r(s_t, a_t) \right],
	\end{equation}
	To refine the decision-making strategy, the actor network is updated by minimizing the gradient of the performance metric \( J(\phi) \), as follows:
	\begin{equation}
		\nabla_{\phi} J(\phi) = \mathbb{E} \left[ \nabla_{a_t} Q_{\theta_1}(s_t, a_t) \bigg|_{a_t=\pi_{\phi}(s_t)} \nabla_{\phi} \pi_{\phi}(s_t) \right].
	\end{equation}
	Concurrently, the critic network is updated to minimize the error between its predictions and the target values \( Y_t \), as specified by the loss function:
	\begin{equation}
		Y_t = r_t + \gamma Q_{\theta_i'}(s_{t+1}, \pi_{\phi'}(s_{t+1}) + {\xi}).
		\label{eq:target}
	\end{equation}
	The proposed approach is encapsulated in Algorithm \ref{algorithm}.

\RestyleAlgo{ruled}
\SetNlSty{textbf}{}{:} 
\begin{algorithm}
    \SetAlgoLined
    \textbf{Initialize:} Experience replay memory \( M \), mini-batch size \( H \), LSTM-based actor \( \pi_{\phi} \) and critic \( Q_{\theta} \) with random parameters. Construct the target networks by setting \( \phi' \gets \phi \) and \( \theta' \gets \theta \).
    
  \textbf{Configuration:}  
Define the episode length \( T \) and the maximum number of episodes \( E \).\\
    
    \For{each episode \( e : E \)}{
       Initialize exploration noise \( \xi \) and state \( s_0 \); \\
        
        \For{each time step \( t \in \{1, \dots, T\} \)}{
            Observe state \( s_t \); \\
            Select an action \( a_t = \pi_{\phi}(s_t) + \xi \), and apply any required reshaping;\\
            Compute reward \( r_t \) based on Eq. (17); \\
            Observe next state \( s_{t+1} \); \\
            Store transition \( (s_t, a_t, r_t, s_{t+1}) \) in \( M \);
        }
        
        Randomly sample a mini-batch of \( H \) experiences from the buffer \( M \);\\
        Compute target \( Y_t \) from Eq. \eqref{eq:target}; \\
        Update actor and critic using Adam optimizer; \\
        Soft-update target networks with blending coefficient \( \tau \in [0, 1] \):
        \[\phi' \leftarrow \tau \phi + (1 - \tau)\phi', \quad \theta' \leftarrow \tau \theta + (1 - \tau)\theta'\]
    }
    \caption{The LSTM-DDPG Algorithm}
    \label{algorithm}
\end{algorithm}

\subsection{Computational Complexity Analysis}

The computational complexity of the proposed algorithm comprises two main components: the action selection process and the training process \cite{222222}. The network architecture consists of an actor network and a critic network, each comprising $\mathcal{U}$ hidden layers with $\mathcal{L}$ neurons per layer, resulting in a computational complexity for action selection that is determined by the size of the consecutive layers, specifically $\mathcal{J} \times (|S| + |A|) \times \mathcal{L}$ for the input and first layer, $\mathcal{L}^2$ for the successive hidden layers, and $\mathcal{L} \times |A|$ for the output layer, where $\mathcal{J}$ represents the length of the previous trajectory, $|S|$ denotes the dimension of the state space, and $|A|$ represents the action dimension.
	
	The computational complexity of the critic network in the proposed LSTM-DDPG algorithm is similarly determined by the size of the consecutive layers, with specific complexities of \(\mathcal{J} \times (|S| + 2 \times |A|) \times \mathcal{L}\) for the input and first layer, \(\mathcal{L}^2\) for the successive hidden layers, and \(\mathcal{L} \times |A|\) for the final connection, where \(|S|\) and \(|A|\) denote the dimensions of the agent state and action spaces, respectively, thus resulting in a total computational complexity for action selection of \(\mathcal{O}(\mathcal{L}^2)\), indicating a quadratic relationship between the number of neurons and the computational burden.
	The training process of the LSTM-DDPG algorithm is characterized by a computational complexity primarily determined by the number of network edges, calculated as $I \times C + C^2 + C \times O$, where $I$ is the input size, $C$ is the number of neurons, and $O$ is the output size \cite{222222}. For the actor and critic networks, this complexity can be further refined to $H |S| \mathcal{L} + H \mathcal{L}^2 + H \mathcal{L} |A|$ and $(H (|S| + |A|) \mathcal{L} + H \mathcal{L}^2 + H \mathcal{L})$, respectively, where $H$ denotes the batch size. Consequently, the overall training complexity for the LSTM-DDPG method is $\mathcal{O}(H \mathcal{L}^2)$. 

Compared to standard DRL methods such as soft actor-critic (SAC)~\cite{333333} and DDPG~\cite{222222}, which typically exhibit a training complexity of
$\mathcal{O}\left( \left( \sum_{N=1}^{N_l} C_N C_{N-1} \right) H \mathcal{N}_e \right)$,
the proposed method reduces computational cost by exploiting temporal dependencies via its LSTM architecture. Furthermore, in comparison to proximal policy optimization (PPO) \cite{444444}, which relies on surrogate loss functions and multiple epochs per training step, LSTM-DDPG achieves a better complexity-performance trade-off in continuous and temporally correlated environments.

	\section{Simulation Results}

This section presents numerical evaluations to demonstrate the performance enhancement achieved by integrating FA arrays with the proposed LSTM-DDPG algorithm. The simulation parameters are initialized as follows~\cite{saeidGlobecom, 10354003, 11489526, 113424121}: the distances between the NOMA users and the AP are independently and uniformly distributed within the range $[20, 40]$ meters, while those between the AirFL users and the AP are uniformly distributed within $[40, 100]$ meters. The AoA for all users is assumed to follow a uniform distribution over $[-\pi/2, \pi/2]$ radians. For the FA array configuration, the parameters are set as $X_0 = 0.5\lambda$, $X = 8\lambda$, and the number of FAs is 5. The Rician factor is \(\kappa_r = 7\), the path loss constants are \(A_L = A_N = 21.98\) dB, and the path loss exponents are \(\alpha_L = 2.09\) for LOS and \(\alpha_N = 3.67\) for NLOS conditions. For simplification, \(\lambda\) is set to 1. The LSTM-DDPG algorithm is trained with a learning rate of 0.0005, a batch size of 64, a replay buffer size of $10^4$, a soft update parameter of 0.001, and a discount factor of 0.9.

  Performance is compared between the FA system and fix position antenna (FPA) using a predetermined location vector \(\boldsymbol{x} = \left[ \frac{X}{L+1}, \ldots, \frac{LX}{L+1} \right]^T\). For the FPA benchmark, the antenna positions are fixed at uniformly spaced locations along the array, while the same LSTM-DDPG framework is used to optimize beamforming and power allocation. This ensures a consistent comparison with the proposed FA-based system. Furthermore, we assess the proposed algorithm against the SAC \cite{333333}, PPO \cite{444444}, and the standard DDPG algorithm \cite{222222}. To assess the learning performance, the average reward over 100 episodes is calculated at each episode \( e \) as \({R_{\text{avg}}(e)} = \frac{1}{100} \sum_{i=e-100}^{e} R_i\), where \( R_i \) denotes the mean reward obtained in episode \( i \). Table I provides a summary of the simulation parameters.

\begin{table}[htbp]
	\caption{System Parameters \cite{saeidGlobecom}, \cite{10354003}.}
	\label{sim}
	\centering
	\begin{tabular}{|c|c|c|}
		\hline
		\textbf{Par.} & \textbf{Description} & \textbf{Value} \\
		\hline
		$B$ & Bandwidth & 1 MHz \\
        \hline
		$N$ & Number of NOMA users & 3 \\
        \hline
		$K$ & Number of AirFL users & 5 \\
		\hline
		$L$ & Number of FAs & 6 \\
		\hline
		$\sigma^2$ & Totle noise power over $B$ & $-114$ dBm \\
		\hline
		$d_N$ & NOMA user–AP distance range & $[20, 40]$ m \\
        \hline
		$d_K$ & AirFL user-AP distance range & $[40, 100]$ m \\
		\hline
		$\phi_{i}$ & AoA range & $[-\pi/2, \pi/2]$ rad \\
		\hline
		$L_0$ & Minimum FA separation & $0.5\lambda$ \\
		\hline
		$[0, L]$ & FA interval distance & $[0, 8\lambda]$ \\
		\hline
		$P^{\text{max}}$ & Maximum transmit power & $36$ dBm \\
		\hline
		$\mathbf{h}_i[\mathbf{x}]$ & Channel gain & Rician \\
		\hline
		$\alpha_L$ & Path loss exponent for LOS links & $2.09$ \\

\hline
		$\alpha_N$ & Path loss exponent for NLOS links & $3.67$ \\
        
		\hline
		$\kappa_r$ & Rician factor & $7$ \\
		\hline
		$A_L, A_N$ & Path loss at reference distance & $21.98$ dB \\
		\hline
		$R_n^{min}$ & Minimum rate required for the $n$-th user & 1 Mbps \\
\hline
			$r_p $ & Penalty factor for bit rate requirement & -1 \\ 
		\hline
		$\mathcal{N}_e$ & Episodes & $6000$ \\
		\hline
		$\tau$ & Target network update period & $0.001$ \\
        \hline
		$H$ & Batch size & $64$ \\
		\hline
		$\mathcal{U}$ & Number of hidden layers & $2$ \\
		\hline
		$-$ & Actor learning rate & $0.0005$ \\
		\hline
		$-$ & Critic learning rate & $0.0001$ \\
		\hline
		$\gamma$ & Discount factor & $0.9$ \\
		\hline
		$M$ & Replay buffer size & $10000$ \\
		\hline
		$-$ & Input/Hidden layer neurons & $200$ \\
		\hline
		$-$ & Output layer neurons & $10$ \\
		\hline
		$-$ & Input/Hidden activation & ReLU \\
		\hline
		$-$ & Output activation & Softmax \\
		\hline
		$-$ & Communication rounds & $100$ \\
		\hline
	\end{tabular}
\end{table}

	Fig. 2 presents the convergence behavior of different DRL algorithms in terms of the average reward achieved over training episodes. The solid curves correspond to the average accumulated rewards, while the shaded regions depict the standard deviation across multiple runs, thereby reflecting the stability and consistency of each method. The results indicate that the LSTM-DDPG algorithm consistently achieves the highest average rewards throughout the training process. Moreover, it exhibits significantly lower variability compared to the baseline methods. These findings underscore the superior performance and robustness of LSTM-DDPG in dynamic environments.

	\begin{figure}
	\centering
	\includegraphics[width=0.85\linewidth, height=5 cm]{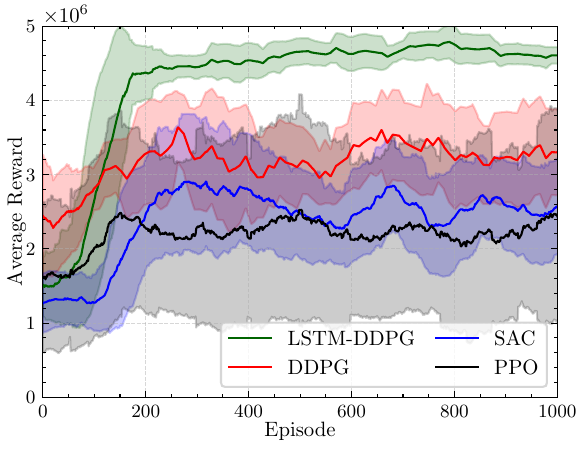}
	\caption{Convergence analysis of DRL algorithms based on average reward per training episode.}
	\label{fig:episode}
\end{figure}
	

Fig.~3 illustrates the impact of various system uncertainties on the hybrid rate performance of the proposed algorithms with different numbers of FAs. Specifically, Fig.~3(a) presents the performance under imperfect SIC, quantified by the uncertainty parameter $\epsilon_b$, while Fig.~3(b) shows the effect of channel estimation errors, where the error variance is denoted by $\sigma_h^2$ and assumed to be identical for all users (i.e., $\sigma_{h,i}^2 = \sigma_h^2, , \forall i$). In both scenarios, the hybrid rate of NOMA and AirFL users improves with increasing number of antennas $L$, although the rate of improvement gradually saturates due to the fixed length of the input signal vector $X$. Additionally, the flexibility offered by FA systems in spatial adaptation leads to consistently superior performance compared to FPA systems. It is important to note that the hybrid rate improvement tends to saturate as $L$ increases. This behavior is primarily attributed to physical-layer limitations. Specifically, since the overall array length $X$ is fixed, increasing $L$ reduces the inter-element spacing, which introduces higher spatial correlation and limits the additional DoF that can be exploited for beamforming. In addition, the total transmit power is constrained by $P^{\text{max}}$, restricting further improvement in the received signal quality as $L$ grows.

\begin{figure}
    \centering
    \subfloat[]{%
        \includegraphics[width=0.85\linewidth, height=5 cm]{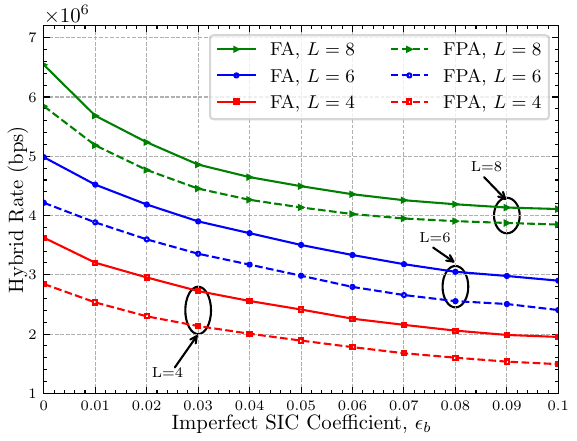}%
        \label{fig:hybrid_a}
    }\\
    \vspace{0.05mm}
    \subfloat[]{%
        \includegraphics[width=0.85\linewidth, height=5 cm]{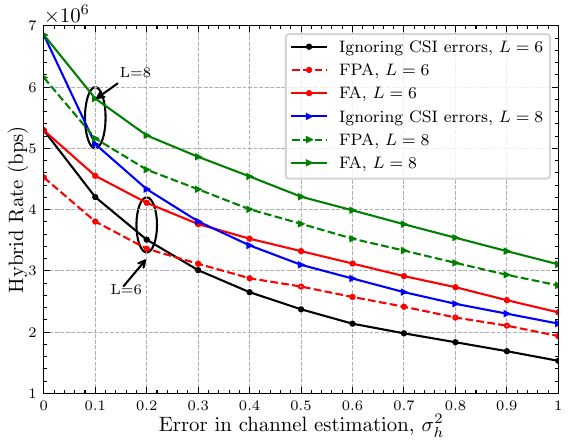}%
        \label{fig:hybrid_b}
    }
    \caption{Hybrid rate performance for different numbers of FAs under two types of network uncertainty: (a) imperfect SIC coefficient \(\epsilon_b\); (b) channel estimation error variance \(\sigma_h^2\).}
    \label{fig:hybrid_rate}
\end{figure}

Fig.~4 illustrates the impact of the total number of AirFL users, denoted by $K$, on the hybrid rate under varying levels of SIC uncertainty $\epsilon_b$ and channel estimation error variance $\sigma_h^2$. It can be observed that the hybrid rate increases consistently with $K$ across all scenarios. This improvement is attributed to the enhanced data aggregation capabilities and the increased diversity gain introduced by a larger number of users. However, the growth rate may taper off under severe uncertainty conditions, particularly when $\epsilon_b$ or $\sigma_h^2$ become large, due to limitations in signal separation and channel reliability. These results highlight the system’s scalability and its robustness in the presence of practical impairments.

\begin{figure}
    \centering
    \subfloat[]{%
        \includegraphics[width=0.85\linewidth, height=5 cm]{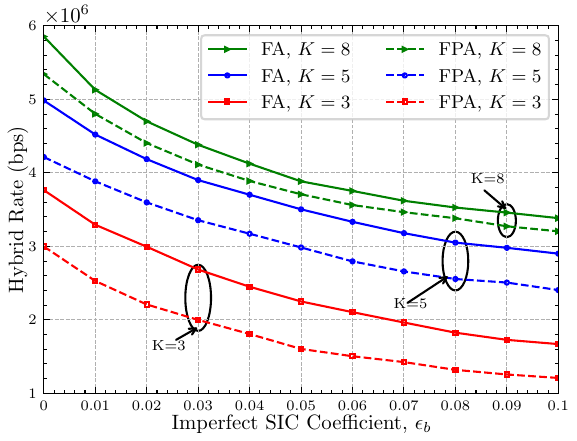}%
        \label{fig:hybrid_a}
    }\\
    \vspace{0.05mm}
    \subfloat[]{%
        \includegraphics[width=0.85\linewidth, height=5 cm]{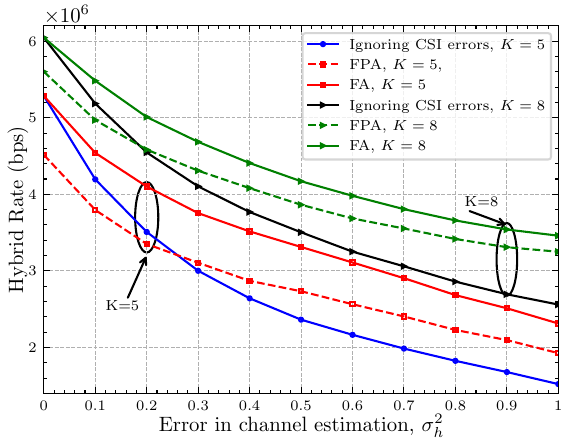}%
        \label{fig:hybrid_b}
    }
    \caption{Hybrid rate performance for different numbers of AirFL users ($K$) under two types of network uncertainty: (a) imperfect SIC coefficient \(\epsilon_b\); (b) channel estimation error variance \(\sigma_h^2\).}
    \label{fig:hybrid_rate}
\end{figure}

\subsection{Experimental Setup and Learning Performance Analysis}

To assess the effectiveness of the proposed FA-assisted hybrid network in realistic conditions affected by channel imperfections and residual interference, comprehensive image classification experiments were carried out using the MNIST dataset.

The evaluation framework considers both independent and identically distributed (IID) and non-IID data distributions to rigorously examine the learning performance of a convolutional neural network (CNN) model under different data partitioning scenarios.

For all experiments, the MNIST dataset is divided into training and testing sets, allocating 90\% of the data for training and the remaining 10\% for testing. Two distinct data distribution strategies are implemented:

\begin{enumerate}
\item \textbf{IID Setting:} Training samples are uniformly shuffled and evenly distributed across all clients, ensuring each client receives an identical distribution of all digit classes.

\item \textbf{Non-IID Setting:} Training samples are sorted by digit labels, and each client is randomly assigned data from only three or four distinct digit classes. This creates an unbalanced distribution where each FL participant possesses data from only a subset of possible digit classes, thereby establishing a challenging non-IID environment.
\end{enumerate}

Each local model is implemented as a feedforward neural network comprising an input layer with 200 neurons and ReLU activation, followed by a hidden layer of 200 neurons with ReLU activation. The output layer includes neurons equal to the number of digit classes and employs a softmax activation function. Additionally, we considered an ideal FL setting without communication noise and full participation as a benchmark.

As shown in Figs.~5(a) and 5(b), which depict the test accuracy and training loss over 200 communication rounds under both IID and non-IID settings, the ideal noise-free FL scheme achieves optimal convergence in both cases. Among all compared approaches, the proposed algorithm yields performance closest to the ideal FL benchmark in terms of both training loss and test accuracy. This demonstrates the effectiveness of utilizing FAs over FPA. Furthermore, the standard deviation of accuracy over the last 10 communication rounds is reduced by 10\% compared to the FPA case, indicating enhanced stability and reliability achieved through antenna mobility. Figs.~5(c) and 5(d) further demonstrate similar trends under the non-IID setting. However, compared to the IID scenario, learning curves under non-IID conditions display more pronounced oscillations and slower convergence rates due to statistical heterogeneity across client data. Nevertheless, the proposed FAS-assisted strategy still maintains a significant advantage over the FPA configuration, affirming its effectiveness in both homogeneous and heterogeneous data environments.


	\begin{figure*}%
		\centering
		\subfloat[\centering]{{\includegraphics[width=4.45cm, height=4.5cm]{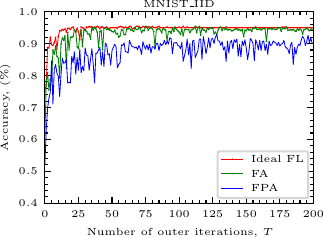} }}%
		\subfloat[\centering]{{\includegraphics[width=4.45cm, height=4.5cm]{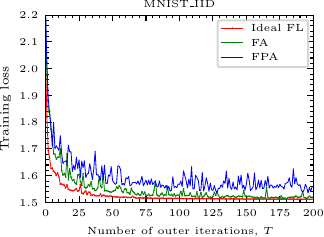} }}%
		\subfloat[\centering]{{\includegraphics[width=4.45cm, height=4.5cm]{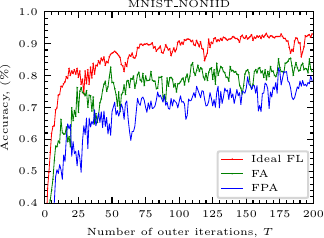} }}%
		\subfloat[\centering]{{\includegraphics[width=4.45cm, height=4.5cm]{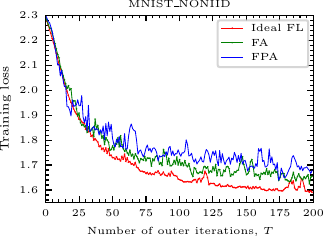} }}%
		\caption{Learning performance over communication rounds on the MNIST dataset: (a)–(b) under IID data distribution; (c)–(d) under non-IID data distribution.}%
		\label{fig:example}%
	\end{figure*}

	\section{Conclusion}

	This paper introduced a framework integrating FA arrays into a hybrid network supporting both AirFL and NOMA users. By leveraging the reconfigurability of FA arrays, the system dynamically adapts to varying channel conditions, thereby mitigating interference and enhancing the overall hybrid rate performance. The study incorporated practical challenges such as imperfect CSI and residual interference from imperfect SIC. To address these issues, a robust joint optimization problem was formulated to minimize the aggregation error while ensuring reliable communication. The problem's inherent non-convexity, due to the coupling between AirFL and NOMA functionalities and CSI/SIC uncertainties, was tackled using a DRL approach based on LSTM-DDPG. Simulation results validated the effectiveness of the proposed FA-assisted strategy, showing significant improvements in hybrid rate performance compared to conventional fixed-antenna schemes, especially in environments with channel uncertainty.

	\bibliographystyle{IEEEtran}
	\bibliography{FA_OTA_FL}

	\vfill
	
\end{document}